\documentstyle[12pt,hyperref]{article}
\newcommand{\be}[1]{ \begin{equation}\label{#1} }
\newcommand{\ee}{\end{equation}}
\newcommand{\bea}[1]{\begin{eqnarray}\label{#1} }
\newcommand{\eea}{\end{eqnarray}}
\newcommand{\eq}[1]{(\ref{#1})}

\newcommand{\II}{{\cal I}}

\newcommand{\FF}{{\cal F}}
\newcommand{\NN}{{\cal N}}
\newcommand{\p}{\partial}
\newcommand{\wt}{\tilde}
\def\ZZZ{{\hskip-3pt\hbox{ Z\kern-1.6mm Z}}}
\def\zzz{{\hskip-3pt\hbox{ z\kern-1mm z}}}

\def\ZZZ{{\hbox{Z\kern-1.6mm Z}}}
\def\zzz{{\hbox{z\kern-1mm z}}}

\newcommand{\vt}{\vartheta}

\newcommand{\ws}{{\wt\sigma}}
\newcommand{\wrh}{{\wt\rho}}
\newcommand{\wv}{{\wt v}}

\newcommand{\dum}{\tau}

\newcommand{\KK}{{\cal K}}

\newcommand{\OO}{{\cal O}}

\newcommand{\EE}{{\cal E}}
\newcommand{\LL}{{\cal L}}

\newcommand{\wh}{\widehat}

\renewcommand{\wh}{\hat}

\newcommand{\ben}{\begin{eqnarray}\displaystyle}
\newcommand{\een}{\end{eqnarray}}
\newcommand{\refb}[1]{(\ref{#1})}
\newcommand{\sectiono}[1]{\section{#1}\setcounter{equation}{0}}

\def\one{{\hbox{ 1\kern-.8mm l}}}
\def\zero{{\hbox{ 0\kern-1.5mm 0}}}

\begin{document}

{}~
{}~

 \hfill\vbox{\hbox{hep-th/0607155}
}\break

\vskip .6cm

{\baselineskip20pt
\begin{center}
{\Large \bf
Dyon Spectrum in $\NN=4$ Supersymmetric Type II String
Theories
}
\end{center} }

\vskip .6cm
\medskip

\vspace*{4.0ex}

\centerline{\large \rm
Justin R. David, Dileep P. Jatkar and Ashoke Sen}

\vspace*{4.0ex}

\centerline{\large \it Harish-Chandra Research Institute}

\centerline{\large \it  Chhatnag Road, Jhusi,
Allahabad 211019, INDIA}

\vspace*{1.0ex}

\centerline{\it E-mail: 
justin,dileep,sen@mri.ernet.in,
ashoke.sen@cern.ch}

\vspace*{5.0ex}

\vskip 1in

\centerline{\bf Abstract} \bigskip

We compute the spectrum of quarter BPS dyons in freely acting
$\ZZZ_2$ and $\ZZZ_3$ orbifolds of type II string theory
compactified on a six dimensional torus. For large
charges the result for statistical entropy computed from the
degeneracy formula agrees with the corresponding black hole
entropy to first non-leading order after taking into account 
corrections due to the
curvature squared terms in the effective action. The result is
significant since in these theories the entropy of a small black
hole, computed using the curvature squared corrections to the
effective action, fails to reproduce the statistical entropy
associated with elementary string states.

\vfill \eject

\baselineskip=18pt

\tableofcontents

\sectiono{Introduction and Summary} \label{s5}

By now there is a reasonably good understanding of the spectrum
of 1/4 BPS dyons in a class of $\NN=4$ supersymmetric string
theories in four dimensons
\cite{9607026,0412287,0505094,0506249,0508174,0510147,
0602254,0603066,0605210}. 
These 
include heterotic string theory on a torus
as well as a class of CHL 
models\cite{CHL,CP,9507027,9507050,9508144,9508154} 
obtained by $\ZZZ_N$ orbifolding
of toroidally compactified heterotic string theory. Dual description
of these theories involve type IIA string theory compactified on
$K3\times T^2$ and appropriate
$\ZZZ_N$ orbifolds of this theory. In each example studied
so far, the statistical entropy computed by taking the logarithm of the
degeneracy of states agrees with the entropy of the corresponding
black hole for large charges, not only in the leading order but also
in the first non-leading order\cite{0412287,0510147,0605210}. 
On the 
black hole side this requires
inclusion of four derivative
terms in the effective action, and use of Wald's generalized formula
for the black hole entropy in the presence of higher derivative
corrections\cite{9307038,9312023,9403028,9502009}.

In this paper we extend this analysis to yet another $\NN=4$
supersymmetric string theory, obtained by taking a freely
acting $\ZZZ_2$ 
orbifold of
type IIA string theory compactified on a six torus $T^6$. The
orbifold group involves reflection of four coordinates of the
torus together with half unit of shift along a fifth direction on
the torus. There is a dual description of this model, also as an
orbifold of type IIA string theory on $T^6$, but now the orbifold
group involves half unit of shift along one coordinate of the
torus together with a $(-1)^{F_L}$ transformation where $F_L$
is the contribution to the space-time fermion number from
the left-moving sector of
the string world-sheet\cite{9508064}. 
Although in many respects this model
has very similar properties to the $\NN=4$ supersymmetric
heterotic string compactification studied earlier, there is one
important difference. Unlike in the $\NN=4$ 
theories coming from heterotic
string compactification, in the present model 
the entropy of a small black
hole representing an elementary string state fails to reproduce
the statistical entropy associated with elementary string
states\cite{0504005,0507014}. This
makes 
it important to test if the statistical entropy of dyons agrees
with the black hole entropy.

We follow the procedure of \cite{0605210} to compute the degeneracy
of a class of dyons in this theory. The result may be summarized as
follows. Let us denote by $Q_e$ and $Q_m$ the electric and magnetic
charge vectors of a state in the second 
description of the theory where the
orbifold group involves a $(-1)^{F_L}$ transformation, and by
$a\cdot b$ the T-duality invariant
inner products between two such charge vectors $a$ and $b$. Then
the degeneracy $d(Q_e,Q_m)$ of a class of 1/4 BPS dyonic states
are given by
\bea{e1int}
d(Q_e, Q_m) &=&   -{1\over 2^9}\, 
\int_{C} d\wt \rho d\wt\sigma d\wt v
{1\over \wt\Phi(\wt \rho, \wt \sigma, \wt v)}
\exp\left[ -i\pi (2\wt \rho Q_e^2 + \wt \sigma Q_m^2/2 +
2\wt v Q_e\cdot Q_m)\right]\, , \nonumber \\
\een
where $Q_e^2\equiv Q_e\cdot Q_e$, $Q_m^2\equiv Q_m\cdot Q_m$,
$\wt\Phi$ is a function to be specified
below, and $C$ is a three real dimensional subspace of the
three complex dimensional space labelled by $(\wt\rho,\ws,\wv)$,
given by
\bea{ep2int}
Im\, \wt \rho=M_1, \quad Im \, \wt\sigma = M_2, \quad
Im \, \wt v = M_3, \nonumber \\
 0\le Re\, \wt\rho\le 1, \quad
0\le Re\, \wt\sigma\le 2, \quad 0\le Re\,  \wt v\le 1\, ,
\een
$M_1$, $M_2$ and $M_3$ being fixed large positive numbers.
The function $\wt\Phi$ is given by
\be{edefwphiint}
\wt \Phi(\wt \rho,\wt \sigma,\wt v ) = -{1\over 2^8}\,
e^{2\pi i (\wt\rho
+ \wt v)}
 \prod_{r=0}^{1}
\prod_{k'\in \zzz+{r\over 2},l,j\in \zzz\atop k',l\ge 0, j<0 \, {\rm for}
\, k'=l=0}
\left( 1 - e^{2\pi i (\wt \sigma k'   + \wt \rho l + \wt v j)}\right)^{
\sum_{s=0}^{1} (-1)^{s l} c^{(r,s)}(4lk' - j^2)}\, ,
 \ee
 where the coefficients $c^{(r,s)}(4lk'-j^2)$ are 
 given as follows. Let us 
denote by $\wt g$ a transformation that changes the
sign of all the coordinates of a four torus $T^4$, and
 consider
 a (4,4) superconformal
field theory (SCFT) with target space $T^4$. We now take an
orbifold of this theory by the 
$\ZZZ_2$ group 
generated by $\wt g$, and
define
 \be{esi4aint}
F^{(r,s)}(\tau,z) \equiv {1\over 2} Tr_{RR;\wt g^r} \left(\wt g^s
(-1)^{F_L+F_R}
e^{2\pi i \tau L_0} e^{2\pi i {\cal J} z}\right), \qquad r,s=0,1\, ,
 \ee
where
$Tr$ denotes trace over all the Ramond-Ramond (RR) sector 
states twisted by $\wt g^r$ in this SCFT 
before we project on to $\wt g$ invariant
states. $F_L$ and $F_R$ denote the world-sheet fermion 
numbers\footnote{For the world-volume theory on the D-branes
the world-volume fermion number coincides with the space-time
fermion number. For describing elementary string states we shall
mostly 
use light-cone gauge Green-Schwarz formalism where again the
world-sheet fermion number coincides with the space-time
fermion number. Thus throughout this paper
there will be no distinction between
world-sheet and space-time fermion numbers.}
associated with left and right chiral fermions in this SCFT, and 
${\cal J}/2$ is the  generator of the $U(1)_L$ subgroup of the
$SU(2)_L\times SU(2)_R$
R-symmetry group of this conformal field 
theory. One finds that $F^{(r,s)}(\tau,z)$ has expansion of the form
\be{enewint}
F^{(r,s)}(\tau,z) =\sum_{b\in\zzz, n} c^{(r,s)}(4n -b^2)
e^{2\pi i n\tau + 2\pi i bz}\, .
\ee
This defines the coefficients $c^{(r,s)}(u)$.

The explicit forms of $F^{(r,s)}(\tau,z)$ are as follows
\bea{valindint}
F^{(0,0)}(\tau, z) &=& 0\nonumber \\
F^{(0,1)}(\tau, z) &=& 8  \,
\frac{\vartheta_2(\tau, z)^2}{\vartheta_2(\tau, 0)^2} \nonumber \\
F^{(1,0)}(\tau, z) &=& 8 \,
\frac{\vartheta_4(\tau, z)^2}{\vartheta_4(\tau, 0)^2} \nonumber \\
F^{(1,1)}(\tau, z) &=& 8 \, \frac{\vartheta_3(\tau, z)^2}
{\vartheta_3(\tau, 0)^2}\, .
\eea
{}From this one can calculate the 
coefficients $c^{(r,s)}(u)$ explicitly.

As in the case of CHL models,
the function $\wt\Phi(\wt \rho,\wt \sigma,\wt v)$ turns out to be a
modular form of weight 2 under a certain subgroup of the Siegel
modular group of genus two Riemann surfaces. Using this fact one
can prove that the degeneracy formula \refb{e1int} is invariant under
the S-duality group $\Gamma_1(2)$ of the theory. 

Using \refb{e1int}
one can also compute the statistical entropy of the dyon for 
large charges following the general strategy outlined in
\cite{0412287,0510147,0605210} 
and compare it with the entropy of the
corresponding black hole. It turns out that up to order $Q^0$
both the statistical entropy and the black hole entropy are obtained
by extremizing the function
\be{blackint}
{\pi\over 2\dum_{2}} \, |Q_e+\dum Q_m|^2
- \ln \wt f (\dum) -\ln \wt f(-\bar\dum)
- 4 \ln (2\dum_{2}) + \hbox{constant} + \OO(Q^{-2})\, ,
\ee
with respect to the real and imaginary parts of $\dum=\dum_1
+i\dum_2$. Here
\be{e10int}
\wt f(\dum) = \eta(\dum)^{16} / \eta(2\dum)^8\, .
\ee
Thus we see that to this order the black hole entropy agrees with
the statistical entropy. The result is significant in light of the fact that
the same four derivative corrections to the effective action fail
to reproduce the statistical entropy of elementary string states in this
theory, essentially due to the fact that these corrections vanish at the
tree level.

These results can also be generalized to a freely acting $\ZZZ_3$
orbifold of type II string theory compactified on a six
dimensional torus. For brevity we shall not give the results here,
but a summary of the results can be found in section \ref{sz3}.

The rest of the paper is organised as follows. In section \ref{s0} we
describe the theory under consideration in different duality
frames, and
also describe the dyon configuration that we shall analyze in this
paper. In section \ref{s2} we count the degeneracy of a class
of 1/4 BPS
dyonic states with a given set of charges, and reproduce eq.\refb{e1int}.
In section \ref{s3} we use the techniques developed in
\cite{0602254} to show that $\wt\Phi$ transforms as a modular form
under a subgroup of $Sp(2,\ZZZ)$. This in turn proves the
S-duality invariance of \refb{e1int}. In section \ref{s1} we analyze
the behaviour of the statistical entropy computed from \refb{e1int}
for large charges and show that it agrees with the black hole
entropy up to first non-leading order. Section \ref{sz3}
contains a summary of the results for the $\ZZZ_3$ orbifold
theory.

Since most of the analysis in this paper is identical to that in 
\cite{0602254,0605210} we often skip the details of the calculation
and quote the final result. For details of the calculation the reader
should consult the original references.

\sectiono{The Dyon Configuration} \label{s0}

In this section we shall describe  the model
under consideration and its various dual descriptions which
will be relevant for our analysis. The analysis is based on the
connection between four and five dimensional
black holes discussed in \cite{0503217,0505094,0605210}.

\begin{enumerate}

\item We begin with type IIB string theory
compactified on a six torus $T^4\times S^1 \times \wt S^1$, 
and take a system containing $Q_5$ D5-branes
wrapped on $T^4\times S^1$, $Q_1$ D1-branes wrapped on
$S^1$, $-n$ units of momentum along $S^1$, $J$ units of
momentum along $\wt S^1$ and a Kaluza-Klein monopole
associated with the compact circle $\wt S^1$.
For definiteness we shall label $S^1$ and $\wt S^1$ by coordinates
with period $2\pi$.
Let us denote
the coordinates of $T^4$ by $x^6, x^7, x^8, x^9$, and the
coordinates of $\wt S^1$,  $S^1$
by $x^4, x^5$.   We then take an orbifold of this system
by a $\ZZZ_2$ transformation generated by
\be{orb1}
g:  (x^4, x^5, x^6, x^7, x^8, x^9) \rightarrow
( x^4, x^5+\pi, -x^6, -x^7, -x^8, -x^9)\, .
\ee
We shall denote by $\wt g$ the part of $g$
that acts on $T^4$, \i.e.
\be{orb1a}
\wt g:  (x^4, x^5, x^6, x^7, x^8, x^9) \rightarrow
( x^4, x^5, -x^6, -x^7, -x^8, -x^9)\, .
\ee
We shall call this the first description of the system.  

\item We now make an S-duality transformation on this system to 
get type IIB string 
theory on
$T^4\times S^1\times \wt S^1/\ZZZ_2$ with
$Q_5$ NS5-branes on $T^4\times S^1$, $Q_1$ units of fundamental 
string winding charge along $S^1$, 
$-n$
units of momentum along $S^1$, $J$ units of momentum
along $\wt S^1$, and a Kaluza-Klein monopole associated with 
$\wt
S^1$ compactification. Under this duality the generators $g$ and
$\wt g$ remain unchanged.

\item
Next make an $R\to 1/R$ duality transformation 
along $\wt S^1$ to 
convert
the theory into type IIA  string theory on 
$T^4\times S^1\times \wh S^1/\ZZZ_2$ with
$Q_5$ Kaluza-Klein monopoles associated with
$\wh S^1$ compactification, $Q_1$ units of 
fundamental string winding charge
along 
$S^1$,
$-n$ units of momentum along $S^1$, $J$ units of 
fundamental string winding charge
along $\wh S^1$, and a single 
NS5-brane wrapped on
$T^4\times S^1$. Here $\wh S^1$ 
denotes the dual circle of $\wt S^1$. Again the generators
$g$ and $\wt g$
remain unchanged under this duality transformation.

\item
Finally using the string-string 
self-duality described in \cite{9508064} we can 
relate this to a type IIA
string theory on $\wh T^4\times S^1\times \wh S^1/\ZZZ_2'$,
where the generator of $\ZZZ_2'$ involves half unit of shift along
$S^1$ together with a $(-1)^{F_L}$ transformation where $F_L$
denotes the contribution to the space-time fermion number from the
left-moving sector of the string world-sheet.
The action of this duality on various states is similar to that
of string-string duality relating type IIA string theory on K3
and heterotic string theory on 
$T^4$.
The final system consists of
$Q_5$ Kaluza-Klein monopoles associated with 
$\wh S^1$ compactification, $Q_1$ units of 
NS5-brane charge along
$\wh T^4\times S^1$, $-n$ units of momentum along $S^1$,
$J$ units of NS5-brane charge along $\wh
T^4\times \wh S^1$, and a single 
fundamental string
wrapped on
$S^1$.
We shall call this description the second description of the system.

\end{enumerate}

Since the second description has only fundamental strings,
NS 5-branes and Kaluza-Klein monopoles, 
we shall use this description  to identify the
various charges as electric or magnetic.
If $-\vec n$ and $\vec w$ denote the 
momentum and winding charges respectively
along $S^1\times \wh S^1$,  and $\vec N$ and $\vec W$
denote the Kaluza-Klein monopole charges and $H$-monopole
charges (NS-5-branes transverse to the circle) along
$S^1\times \wh S^1$, then we can define the T-duality
invariant inner product
\be{etdual}
Q_e^2 = 2\vec n\cdot \vec w, \quad Q_m^2 =2\vec N\cdot \vec W,
\quad Q_e\cdot Q_m = \vec n \cdot \vec N + \vec w\cdot \vec W
\, .
\ee
Thus before the $\ZZZ_2$ modding we had 
${1\over 2} \,
Q_m^2 = Q_1 Q_5 $, ${1\over 2}
Q_e^2 = n$, and $Q_e . Q_m = J$.
In order to get a $\ZZZ_2$ invariant 
configuration so that we can carry out the $\ZZZ_2$ modding, 
we need 
to put periodic boundary conditions on all the 
branes which extend along $S^1$, and take 
$2$ identical copies of all the 
branes 
transverse to $S^1$ and place them at intervals of 
$\pi$ along $S^1$. 
The latter 
set includes the five branes along $\wh
T^4\times \wh S^1$;  we need to take 
$2J$ five branes, divide them into two sets
and place the two sets separated 
by an interval of $\pi$ along $S^1$. 
After orbifolding the 
direction along $S^1$ can be regarded as a 
circle of radius $1/2$, and per unit period along $S^1$ there will
be $J$ five branes transverse to $S^1$. The 
natural unit of momentum along $S^1$ is now $2$, and momentum 
$-n$ along $S^1$ can be regarded as $-n/2$ units of momentum. 
The other 
charges have the same values as in the parent theory.  
Thus we now have
\be{echarges}
{1\over 2}Q_e^2=n/2, \qquad 
{1\over 2} Q_m^2 =Q_1\, Q_5, \qquad
Q_e\cdot Q_m = J\, .
\ee

Before concluding this section we shall make a few remarks about
the supersymmetry and S-duality symmetry
of the theory and also the spectrum of massless states
in the theory. Type II string theory compactified
on torus has 32 supercharges, but the $\ZZZ_2$ orbifolding breaks
half of these supersymmetries. In the first description half of the
supersymmetries from the left-moving sector of the world-sheet
and half of the supersymmetries from the right-moving sector
of the world-sheet are broken. Thus this description is analogous
to type II string theory compactified on $K3\times T^2$. 
In the second
description all the supersymmetries from the left-moving sector
of the world-sheet are broken and all supersymmetries from the
right-moving sector of the world-sheet are preserved. Thus this
situation is analogous to heterotic string theory on $T^6$. As in
\cite{0605210} the dyon system breaks 3/4 of the supersymmetry
generators; hence these are 1/4 BPS states of the theory.

The S-duality symmetry of this theory in the second description
may be analysed by mapping
it to the T-duality symmetry of the theory in the first description.
It is essentially the subgroup of the T-duality symmetry 
$SL(2,\ZZZ)$ of $T^2$ that commutes with half unit of shift along
$S^1$, and is generated by the group of matrices $\pmatrix{a &b
\cr c & d}$ satisfying
\be{descra}
ad-bc=1, \quad a,d\in 1 + 2\ZZZ, \quad c \in 2\ZZZ, \quad b\in \ZZZ\, .
\ee
This defines the group 
$\Gamma_1(2)\equiv\Gamma_0(2)$\cite{9507050}.

The spectrum of massless states may be analyzed 
easily using the second
description of the theory. First of all since the theory has $\NN=4$
supersymmetry, the low energy effective field theory must be
$\NN=4$ supergravity coupled to a set of matter multiplets. Thus
in order to find the spectrum all we need to do is to find the number
of matter multiplets. This in turn is equal to the number of
massless vector fields (the rank of the gauge group) minus six, since
there are six graviphotons. To count the number of massless
vector fields we note that since in the Neveu-Schwarz-Ramond (NSR)
formulation the $\ZZZ_2$ transformation
changes the sign of all the Ramond (R) sector states on the left, 
it projects out all the massless
states (including the gauge fields) originating in the
RR sector. On the other hand since it acts
trivially on the massless NS-NS sector states,
all the 12 gauge fields
in the NS-NS sector coming from the components of the metric
and rank two
anti-symmetric tensor fields along the internal directions of
the torus survive the projection. 
This gives a rank 12 gauge group. Thus we have 
six matter multiplets.

\sectiono{Counting of States of the Dyon} \label{s2}

The description of the system given in the previous section makes
it clear that the system is very similar to the corresponding
system in the $\ZZZ_2$ CHL model analyzed in \cite{0605210} 
with
$K3$ replaced by $T^4$, and the transformation $\wt g$ 
given by \refb{orb1a} rather than a $\ZZZ_2$ involution
in K3. Thus the computation of the degeneracy proceeds in a manner
identical to that in \cite{0605210}. We now
outline the main
steps in this computation.

As shown in \cite{0605210} the final result for the degeneracy
depends only on the combination $Q_1Q_5$; hence we shall
for simplicity consider the $Q_5=1$ case.\footnote{Unlike in
\cite{0605210} where wrapping a D5-brane on $K3$ shifted
the $Q_1$ charge by $-Q_5$, a D5-brane wrapped on $T^4$
does not cause any such shift.}
In the first description of the system 
the quantum numbers  $n$ and $J$
arise from three different sources: the
excitations of the Kaluza-Klein monopole which can carry
certain amount of momentum $-l_0'$ along $S^1$, the  overall 
motion of the D1-D5 system in the background of the
Kaluza-Klein monopole which can carry certain amount of
momentum $-l_0$ along $S^1$ and $j_0$ along $\wt S^1$ and
the  motion of the D1-branes  in the plane of the
D5-brane carrying  total  momentum $-L$ along $S^1$ and $J'$
along $\wt S^1$. Thus we have
\be{esi1}
  l_0'+l_0+
L = n, \qquad j_0+  J'=J\, .
\ee
Let $h(Q_1, n, J)$ denote the number of bosonic minus fermionic
supermultiplets (in the sense described in \cite{0605210})
of the combined system carrying quantum numbers
$Q_1$, $n$, $J$ and let
\be{esi8}
f(\wt \rho,\wt \sigma,\wt v) = \sum_{Q_1, n, J} h(Q_1,n, J)
e^{2\pi i ( \wt \rho n + \wt \sigma Q_1/2 +\wt v J)} \, ,
\ee
denote the partition function of the system. Then 
$f(\wt \rho,\wt \sigma,\wt v)$ is obtained as a product of
three separate partition functions:
\bea{esi9}
f(\wt \rho,\wt \sigma,\wt v ) &=& {1\over 64} 
\, \sum_{Q_1,L,J'} d_{D1}(Q_1,L,J') 
e^{2\pi i ( 
\wt \sigma Q_1 /2 +\wt \rho L + \wt v J')}
\nonumber \\
&& \, \left(\sum_{l_0, j_0} 
d_{CM}(l_0, j_0) e^{2\pi i l_0\wt\rho + 2\pi i
j_0\wt v}\right) \, 
\left(\sum_{l_0'} d_{KK}(l_0') e^{2\pi i l_0' \wt\rho} \right)\, ,
\eea
where $d_{D1}(Q_1,L,J')$ is the degeneracy of $Q_1$ D1-branes
moving in the plane of the D5-brane carrying momenta $(-L,J')$
along $(S^1,\wt S^1)$, $d_{CM}(l_0,j_0)$ is
the  degeneracy associated with the overall
motion of the D1-D5 system
in the background of the Kaluza-Klein monopole carrying
momenta $(-l_0, j_0)$ along $(S^1,\wt S^1)$ and
$d_{KK}(l_0')$ denotes the degeneracy associated with
the excitations of a Kaluza-Klein monopole carrying momentum
$-l_0'$ along $S^1$. The factor of 1/64
in \refb{esi9} accounts for the fact that a single 1/4 BPS
supermultiplet has 64 states.

We begin with the computation of $d_{KK}(l_0')$.
Under the duality that relates the first description to the
second description, a Kaluza-Klein monopole in the first description
gets 
mapped to
a twisted sector fundamental string in the second description, and the
transformation $\wt g$ gets mapped to $\wh g=(-1)^{F_L}$.
Let us consider a (4,4) superconformal field theory describing
type IIA string theory compactified on $T^4\times S^1\times \wh S^1$
in the light-cone gauge Green-Schwarz formalism.
Following the procedure of \cite{0605210} one
finds that
\be{e3.1}
\sum_{l_0'} d_{KK}(l_0') e^{2\pi i l_0' \wt\rho}
= 
Tr'_{\wh g} \left(  (-1)^{F_L} e^{4\pi i \rho L_0'}\right)\, ,
\ee
where  $Tr'_{\wh g}$ denotes trace over states for which
the right-moving oscillators are in their ground state, 
and the left-moving oscillators are
twisted by $\wh g$. We do not impose the requirement of 
$\wh g$ invariance on the states while 
taking the trace\cite{0605210}. 
The factor of $(-1)^{F_L}$ inside the trace accounts for the fact
that we want to count bosonic and fermionic excitations in
the left-moving sector of the world-sheet with weights 1 and $-1$
respectively. This factor was not present in
the corresponding expression in
\cite{0605210} 
since all the left-moving world-sheet oscillators were
bosonic.
The Virasoro generator
$L_0'$ includes the
contribution from all the left moving bosonic and fermionic
oscillators but not from momenta or winding charges
which are set to
some fixed values. Since
in the Green-Schwarz formulation
there are 8 left-moving bosonic oscillators with periodic boundary
condition and 8 left-moving fermionic oscillators with
anti-periodic boundary condition (due to twisting by $\hat g$
under which the fermions are odd) we get
\be{e3.2}
\sum_{l_0'} d_{KK}(l_0') e^{2\pi i l_0' \wt\rho}
= 16 \, e^{-2\pi i \wt\rho}
{\prod_{n=1}^\infty (1 - e^{2\pi i (2n-1)\wt\rho})^8 
\over \prod_{n=1}^\infty (1 - e^{4\pi i n\wt\rho})^8}
= 16\, {\eta(\wt\rho)^8\over \eta(2\wt\rho)^{16}}\, .
\ee
The factor of 16 comes from the fermionic zero mode quantization
in the right-moving sector. The overall factor of 
$e^{-2\pi i \wt\rho}$ reflects the effect of the zero point energy.

Next we compute $d_{CM}(l_0, j_0)$. In this case 
besides the degrees of freedom associated with the motion
of the D1-D5 system transverse to the plane of the D5-brane
as in \cite{0605210},
there is an additional set of degrees of freedom associated with
the Wilson lines  along $T^4$  on the D5-brane\cite{0203048}. 
This
gives rise to four additional bosonic fields together with their
fermionic superpartners living on $S^1$. For the degeneracy 
associated with the dynamics transverse to the plane of the
D5-brane,
not only
the computational procedure but also the results are identical to that
in \cite{0605210} for the $\ZZZ_2$ orbifold case, 
and we get
\bea{eone1}
&& \sum_{l_0, j_0} d_{transverse}(l_0, j_0) e^{2\pi i l_0\wt\rho + 2\pi i
j_0\wt v} = 4 \, e^{-2\pi i \wt v} \,  (1 - e^{-2\pi i \wt v})^{-2}\,
\nonumber \\
&& \qquad \qquad \prod_{n=1}^\infty \left\{
(1 - e^{4\pi i n \wt\rho})^4 \, ( 1 - e^{4\pi i n \wt\rho + 2\pi i
\wt v})^{-2} \,  ( 1 - e^{4\pi i n \wt\rho - 2\pi i
\wt v})^{-2}\right\}\, .
\eea
On the other hand the bosonic fields associated with the
Wilson line along $T^4$ and their fermionic superpartners
are odd under $\wt g$, and hence have anti-periodic
boundary condition along $S^1$. 
Together they describe a (4,4) superconformal field theory with
$SU(2)_L\times SU(2)_R$ R-symmetry, and the quantum number
$j_0$ may be identified with twice the eigenvalue of the $U(1)_L$
generator of $SU(2)_L$\cite{9602065}.
The bosons and the right-moving
fermions are neutral under $SU(2)_L$ and hence do not carry
any $j_0$ quantum number, but the left-moving 
fermions are doublets under the $SU(2)_L$ R-symmetry
group and hence carry $j_0$ quantum numbers 
$\pm 1$.\footnote{Recall that the fermions which are superpartners
of the bosonic fields representing transverse motion of the D-brane
have exactly opposite properties. The left-moving fermions are
neutral under $SU(2)_L$ and the right-moving fermions transform
in the doublet representation of $SU(2)_L$\cite{0605210}.}
Thus we have
\bea{eone2}
\sum_{l_0, j_0} d_{wilson}(l_0, j_0)
e^{2\pi i l_0\wt\rho + 2\pi i
j_0\wt v}
 &=& \prod_{n=1}^\infty \left\{
(1 - e^{2\pi i (2n-1) \wt\rho})^{-4} \, ( 1 - e^{2\pi i (2n-1) 
\wt\rho + 2\pi i
\wt v})^{2} \,  \right.\nonumber \\
&& \left. \qquad ( 1 - e^{2\pi i (2n-1) \wt\rho - 2\pi i
\wt v})^{2}\right\}\, .  
\eea
The partition function associated with $d_{CM}(l_0, j_0)$ is given
by the product of these two contributions:
\bea{eone}
&& \sum_{l_0, j_0} d_{CM}(l_0, j_0) e^{2\pi i l_0\wt\rho + 2\pi i
j_0\wt v} = 4 \, e^{-2\pi i \wt v} \,  (1 - e^{-2\pi i \wt v})^{-2}\,
\nonumber \\
&& \qquad \qquad \prod_{n=1}^\infty \left\{
(1 - e^{4\pi i n \wt\rho})^4 \, ( 1 - e^{4\pi i n \wt\rho + 2\pi i
\wt v})^{-2} \,  ( 1 - e^{4\pi i n \wt\rho - 2\pi i
\wt v})^{-2}\right\}  \nonumber \\
&& \qquad \qquad \prod_{n=1}^\infty \left\{
(1 - e^{2\pi i (2n-1) \wt\rho})^{-4} \, ( 1 - e^{2\pi i (2n-1) 
\wt\rho + 2\pi i
\wt v})^{2} \,    ( 1 - e^{2\pi i (2n-1) \wt\rho - 2\pi i
\wt v})^{2}\right\}\, .  \nonumber \\
\eea

Finally we need to find $d_{D1}(Q_1,L,J')$. Since the analysis
is identical to the one given in \cite{9608096,0605210}, 
we shall only
quote the result. We first define
\be{esi4a}
F^{(r,s)}(\tau,z) \equiv {1\over 2} Tr_{RR;\wt g^r} \left(\wt g^s
(-1)^{F_L+F_R}
e^{2\pi i \tau L_0} e^{2\pi i {\cal J} z}\right), \qquad r,s=0,1\, ,
 \ee
where the trace is taken over all the RR sector 
states twisted by $\wt g^r$ in a (4,4) superconformal
field theory with target space $T^4/\ZZZ_2$, --
with $\ZZZ_2$
generated by $\wt g$,  -- before we project on to $\wt g$ invariant
states. $F_L$ and $F_R$ denote the world-sheet fermion numbers 
associated with left and right chiral fermions, and 
${\cal J}/2$ is the  generator of the $U(1)_L$ subgroup of the
$SU(2)_L\times SU(2)_R$
R-symmetry group of this conformal field 
theory. Explicit computation gives
\bea{valind}
F^{(0,0)}(\tau, z) &=& 0\nonumber \\
F^{(0,1)}(\tau, z) &=& 8  \,
\frac{\vartheta_2(\tau, z)^2}{\vartheta_2(\tau, 0)^2} \nonumber \\
F^{(1,0)}(\tau, z) &=& 8 \,
\frac{\vartheta_4(\tau, z)^2}{\vartheta_4(\tau, 0)^2} \nonumber \\
F^{(1,1)}(\tau, z) &=& 8 \, \frac{\vartheta_3(\tau, z)^2}
{\vartheta_3(\tau, 0)^2}\, .
\eea
These can be rewritten as
\be{ee3}
F^{(r,s)}(\tau, z) = h_0^{(r,s)}(\tau) 
 \, \vartheta_3(2\tau, 2z)
+ h^{(r,s)}_1(\tau) 
\, \vartheta_2(2\tau, 2z) 
\ee
where
\bea{defhlrs}
h_0^{(0,0)}(\tau) &=& 0, \qquad
h_1^{(0,0)}(\tau ) = 0, \cr
h_0^{(0,1)}(\tau) &=& 4 \frac{1}{\vartheta_3(2\tau, 0)}, \qquad
h_1^{(0,1)}(\tau) = 4 \frac{1}{\vartheta_2(2\tau, 0)}, \cr
h_0^{(1,0)}(\tau) &=& 8 \frac{\vartheta_3(2\tau, 0)}
{\vartheta_4(\tau, 0)^2}, 
\quad 
h_1^{(1,0)}(\tau) = -8 \frac{\vartheta_2(2\tau, 0)}
{\vartheta_4(\tau, 0)^2}, 
\cr
h_0^{(1,1)}(\tau) &=& 8 \frac{\vartheta_3(2\tau, 0)}
{\vartheta_3(\tau, 0)^2}, 
\quad 
h_1^{(1,1)}(\tau) = 8 \frac{\vartheta_2(2\tau, 0)}
{\vartheta_3(\tau, 0)^2}.
\eea
We now define the coefficients $c^{(r,s)}(u)$ through the expansions
\be{defcrs}
h_0^{(r,s)} (\tau) = \sum _n c^{(r,s)}(4n) q^n, \qquad
h_1^{(r,s)} (\tau) = \sum_n c^{(r,s)}(4n) q^n\, .
\ee
{}From \eq{defhlrs} we see that in the expansion of 
$h_l^{(r,s)}, n \in\ZZZ - \frac{l}{4}$ for $r=0$ and 
$n\in \frac{1}{2}\ZZZ - \frac{l}{4}$ for $r=1$.  
Thus for given $(r,s)$ the 
$c^{(r,s)}(u)$ defined through the two equations in \refb{defcrs}
have non-overlapping  set of arguments. Substituting \refb{defcrs}
into \refb{ee3} and using the Fourier expansions of $\vt_3(2\tau,2z)$,
$\vt_2(2\tau,2z)$ we get
\be{e1.5copy}
F^{(r,s)}(\tau,z) =\sum_{b\in\zzz, n} c^{(r,s)}(4n -b^2)
e^{2\pi i n\tau + 2\pi i bz}\, .
\ee

Following the analysis of \cite{0605210} one can show that
\be{emulti}
\sum_{Q_1,L,J'} d_{D1}(Q_1,L,J') e^{2\pi i ( 
\wt \sigma Q_1/2 +\wt \rho L + \wt v J')}
= \prod_{w,l,j\in \zzz\atop w>0, l\ge 0}  
\left( 1 - e^{2\pi i (\wt \sigma w / 2 + \wt \rho l + 
\wt v j)}\right)^{-n(w,l,j)}\, ,
\ee
where
\be{esi7.1}
n(w,l,j)  
= 
\sum_{s=0}^{1} (-1)^{sl} c^{(r,s)}(2lw - j^2)\, ,
\qquad \hbox{$r=w$ mod $2$}\, .
\ee

It is now time to put the results together. Using the results
\be{ecrsv}
 c^{(0,0)}(0) = 0\, , \qquad 
c^{(0,0)}(-1) = 0\, , \qquad c^{(0,1)}(0) = 4\, , 
\qquad c^{(0,1)}(-1) = 2\, ,  
\ee
and eqs.\refb{esi9}, \refb{e3.2}, \refb{eone} and \refb{emulti}
we get
\be{esi9ex}
f(\wt \rho,\wt \sigma,\wt v )= e^{-2\pi i (\wt\rho
+ \wt v)}
 \prod_{r=0}^{1}
\prod_{k'\in \zzz+{r\over 2},l,j\in \zzz\atop k',l\ge 0, j<0 \, {\rm for}
\, k'=l=0}
\left( 1 - e^{2\pi i (\wt \sigma k'   + \wt \rho l + \wt v j)}\right)^{
-\sum_{s=0}^{1} (-1)^{sl}\, c^{(r,s)}(4lk' - j^2)}\, .
 \ee
Defining
\be{edefwphi}
\wt \Phi(\wt \rho,\wt \sigma,\wt v ) = -{1\over 2^8}\,
e^{2\pi i (\wt\rho
+ \wt v)}
 \prod_{r=0}^{1}
\prod_{k'\in \zzz+{r\over 2},l,j\in \zzz\atop k',l\ge 0, j<0 \, {\rm for}
\, k'=l=0}
\left( 1 - e^{2\pi i (\wt \sigma k'   + \wt \rho l + \wt v j)}\right)^{
\sum_{s=0}^{1} (-1)^{sl} c^{(r,s)}(4lk' - j^2)}\, ,
 \ee
 we can express \refb{esi9ex} as
 \be{e10.1}
 f(\wt \rho,\wt \sigma,\wt v ) = -{1\over 2^8\,
 \wt \Phi(\wt \rho,\wt \sigma,\wt v )}
 \, .
 \ee
 Using \refb{esi8} and identifying $h(Q_1, n,J)$ with 
 the dyonic degeneracy $d(Q_e,Q_m)$ with
 $Q_e^2 = n $, $Q_m^2=2 Q_1$ and $Q_e\cdot Q_m=J$,
 we get
 \bea{e1pre}
d(Q_e, Q_m) &=&   K
\int_{C} d\wt \rho d\wt\sigma d\wt v
{1\over \wt\Phi(\wt \rho, \wt \sigma, \wt v)}
\exp\left[ -i\pi (2\wt \rho Q_e^2 + \wt \sigma Q_m^2/2 +
2\wt v Q_e\cdot Q_m)\right]\, , \nonumber \\
\een
where
\be{edefk}
K=-{1\over 2^9}\, ,
\ee
 and $C$ denotes the surface
\bea{ep2}
Im\, \wt \rho=M_1, \quad Im \, \wt\sigma = M_2, \quad
Im \, \wt v = M_3, \nonumber \\
 0\le Re\, \wt\rho\le 1, \quad
0\le Re\, \wt\sigma\le 2, \quad 0\le Re\,  \wt v\le 1\, ,
\een
$M_1$, $M_2$, $M_3$ being fixed large positive numbers.

\sectiono{Properties of $\wt\Phi$ from the Threshold Integral}
\label{s3}

In this section we shall  derive various useful properties of
$\wt\Phi$, {\it e.g.} its duality transformation laws and
locations of its zeroes  by following the strategy
described in \cite{0602254,0605210} for CHL models. The 
main idea is to begin with an integral that is manifestly invariant
under a subgroup of the modular group $Sp(2,\ZZZ)$ of genus
two Riemann surface and then express this as a sum of a
holomorphic piece proportional to
$\ln\wt\Phi$, its complex conjugate and a piece that
is neither holomorphic nor anti-holomorphic but has simple
transformation properties under $Sp(2,\ZZZ)$ duality transformation.
This in turn would determine the modular transformation
laws of the holomorphic and the anti-holomorphic
pieces separately.

\subsection{The threshold integral}

We  define as in \cite{0602254}
\bea{ex3}
F_{m_1, m_2, n_1, n_2}(\tau, z) 
&=& \sum_{s=0}^1\, (-1)^{m_1\, s}\, 
F^{(r,s)}(\tau,z)  \quad \hbox{for $m_1,m_2,n_2\in\ZZZ$,
$n_1\in \ZZZ+{r\over 2}$, 
$r=0,1$}\nonumber \\
&\equiv& \sum_b F_{m_1, n_1, m_2, n_2;b}(\tau)
\, e^{2\pi i b z}
\eea
and
\be{ef2}
\II(\wrh ,\ws,\wv ) = \int_{\FF} {d^2\tau\over \tau_2} \,
\sum_{m_1, m_2,  n_2, b\in \zzz\atop
n_1\in{1\over 2}\zzz} q^{p_L^2/2-b^2/4} \bar
q^{p_R^2/2}
F_{m_1, m_2, n_1, n_2;b}(\tau)
\ee
where $\FF$ denotes the fundamental domain of $SL(2,\ZZZ)$ in the
upper half plane, $F^{(r,s)}(\tau,z)$ have been defined in 
\refb{esi4a}, and 
\be{edefq}
q=e^{2\pi i\tau}\, ,
\ee
\bea{e7}
{1\over 2} p_R^2 &=& {1\over 4 \det Im \Omega} |-m_1 \wrh  +
m_2 + n_1 \ws + n_2 (\ws\wrh -\wv ^2) + b \wv |^2, \nonumber \\
{1\over 2} p_L^2
&=& {1\over 2}  p_R^2 + m_1 n_1 + m_2 n_2 + {1\over 4} b^2\, ,
\eea
\be{edefomega}
\Omega=\pmatrix{\wrh  & \wv \cr \wv  & \ws}\, .
\ee
Using \refb{ee3} the integral in \eq{ef2} can  be written as
\be{rwthrint}
{\cal I}(\wrh , \ws, \wv ) = \sum_{l, r, s =0}^1 {\cal I}_{r, s, l} 
\ee
\be{defirsl}
{\cal I}_{r,s,l} = \int_{\FF} \frac{d^2\tau}{\tau_2} 
\sum_{\stackrel{m_1, m_2, n_2 \in \zzz}{n_1\in \zzz+ \frac{r}{2}, 
b\in 2\zzz + l}}
q^{p_L^2/2} \bar q^{ p_R^2/2} (-1)^{m_1s} h_l^{(r,s)}(\tau)\, .
\ee
These integrals can be evaluated following the procedure of
\cite{dixon,9512046,0602254} by separately 
evaluating the contribution
from the zero orbit, the degenerate orbits and the
non-degenerate orbits.  The only difference in the result from that
in \cite{0602254}
arises from the fact that the coefficients $c^{(r,s)}(4n-b^2)$ 
now have
different values. The final result is given by:
\bea{fullint1}
{\cal I}(\wrh ,\ws,\wv ) &=& -2 \ln \bigg[
\kappa ( \det\,{\rm Im}\Omega)^2 \bigg|
\exp( 2\pi i ( \wrh  + \wv ))    \nonumber \\
&\;&  \prod_{r,s=0}^1 
\prod_{\stackrel{(l, b)\in \zzz, k'\in \zzz+ \frac{r}{2}}{ k',l\ge 0,
b<0 \, {\rm for}\, k'=l=0}}
\left\{
( 1- \exp(2\pi i (k'\ws+ l\wrh  + b\wv )))^{(-1)^{ls} c^{(r,s)}( 4k'l -b^2)}
\right\}
\bigg|^2 \bigg] \nonumber \\
&=& -2 \ln \left[2^{16}
\kappa ( \det\,{\rm Im}\Omega)^2 \right] - 2\ln \wt\Phi (\wrh ,\ws,\wv )
- 2\ln \bar{\wt\Phi}(\wrh ,\ws,\wv )
\eea
where $\wt\Phi$ has been defined in \refb{edefwphi} and
\be{defka}
\kappa = \left( \frac{8\pi}{3\sqrt{3}} e^{1-\gamma_E} \right)^2.
\ee
In arriving at \refb{fullint1} we have used
\bea{valcs}
c^{(0,0)}(0)=0, \quad c^{(0,0)}(-1)=0, \quad 
c^{(0,1)}(0) =4, \quad c^{(0,1)}(-1) =2, \cr
c^{(1,0)}(0) =8, \quad c^{(1,0)}(-1) =0, \quad
c^{(1,1)}(0) =8, \quad c^{(1,1)}(-1) =0.
\eea

Another useful integral is
\be{eiprime}
\II'(\wrh ,\ws,\wv ) = \II\left(
\wrh  -{\wv ^2\over \ws},-{1\over \ws},
     {\wv \over \ws}\right)\, .
\ee
By manipulating the expression for $\II(\wrh ,\ws,\wv )$ given in
\refb{ef2} and the duality transformation properties
of $p_L^2$ and $p_R^2$ one can show that\cite{0602254}
\be{rwthrintnew}
{\cal I'}(\wrh , \ws, \wv ) = \sum_{l, r, s =0}^1 {\cal I'}_{r, s, l} 
\ee
\be{defirslnew}
{\cal I'}_{r,s,l} = \int_\FF \frac{d^2\tau}{\tau_2} 
\sum_{\stackrel{m_1, n_1, n_2 \in \zzz}{m_2\in \zzz+ \frac{r}{2}, 
b\in 2\zzz + l}}
q^{p_L^2/2} \bar q^{ p_R^2/2} (-1)^{n_2 s} h_l^{(r,s)}(\tau)\, .
\ee
These integrals may also be analyzed following the procedure
described in \cite{0602254} and the result is
\be{erewritenew}
\II'(\wrh ,\ws,\wv ) =  -2 \ln \left[2^{16}
\kappa ( \det\,{\rm Im}\Omega)^2 \right]
 -2 \ln\Phi(\wrh ,\ws,\wv ) -2 \ln \bar
{\Phi}(\wrh ,\ws,\wv )  \, ,
\ee
where
\bea{defpi10new}
\Phi (\rho ,\sigma,v ) &=& -\exp(2\pi i (
\sigma+ \rho  + v ) ) \nonumber \\
&& \prod_{r,s=0}^1\, \prod_{(k',l,b)\in \zzz\atop 
k',l\ge 0, b<0 \, {\rm for}\, k'=l=0}
\Big\{ 1 - (-1)^r\, 
\exp( 2\pi i ( k' \sigma + l \rho  + b v ) \Big\}^{ c^{(r,s)}(4k'l - b^2)}  \, .
\nonumber \\
\eea
It follows from \refb{fullint1}, \refb{erewritenew} and the relation
\refb{eiprime} between $\II$ and $\II'$ that\footnote{This analysis
does not determine the relative phase between $\Phi$ and $\wt\Phi$.
This can be fixed by comparing the $v\to 0$ (or $\wv\to 0$)
limit of the two sides of eq.\refb{reln}.}
\be{reln}
\Phi(\rho ,\sigma,v ) = \sigma^{-2}\, \wt\Phi\left(
\rho  -{v ^2\over \sigma},-{1\over \sigma},
     {v \over \sigma}\right)\, , \quad
\wt\Phi(\wrh ,\ws,\wv ) = \ws^{-2}\,  \Phi\left(
\wrh  -{\wv ^2\over \ws},-{1\over \ws},
     {\wv \over \ws}\right)\, .
\ee
We shall now use these relations to analyze various properties
of $\wt\Phi(\wrh ,\ws,\wv )$.

\subsection{Duality transformation properties}

Following the same line of argument as in \cite{0602254} for
the $\ZZZ_2$ CHL model,
the original integral $\II(\wrh ,\ws,\wv )$ can be shown to be invariant
under a transformation:
\be{edutrs}
\Omega\to (A\Omega+B)(C\Omega + D)^{-1}\, ,
\ee
if
the matrix $\pmatrix{A & B\cr C & D}$
belongs to a subgroup $\wt G$ of $Sp(2,\ZZZ)$ defined in
\cite{0510147}. 
Using the  invariance of $\II$ under \refb{edutrs}
and the relation 
\refb{fullint1} we see that $\wt\Phi$
is a modular form of weight $2$ under the subgroup 
$\wt G$ of $Sp(2, \ZZZ)$:
\be{emodul}
\wt \Phi\left((A\Omega+B)(C\Omega+D)^{-1}\right)
= \det(C\Omega+D)^2 \wt\Phi(\Omega), \qquad
\pmatrix{A & B\cr C & D}\in \wt G\, .
\ee
Using this result we can now follow the procedure of
\cite{0510147} to establish the  invariance of
$d(Q_e,Q_m)$ under the duality transformation:
\be{try1}
\pmatrix{Q_m/\sqrt 2\cr \sqrt 2 Q_e}
\to \pmatrix{a & b\cr c & d} \pmatrix{Q_m/\sqrt 2\cr \sqrt 2 Q_e}
\ee
with $\pmatrix{a & b\cr c & d} \in\Gamma_1(2)$, \i.e.\ 
\be{descr}
ad-bc=1, \quad a,d\in 1 + 2\ZZZ, \quad c \in 2\ZZZ, \quad b\in \ZZZ\, .
\ee
These transformation laws are somewhat different in appearance from
the standard duality transformation laws discussed {\it e.g.} in
\cite{0510147}. This is due to the fact that the degeneracy formula
\refb{e1pre} is related to the corresponding formula 
in \cite{0510147} by
the transformation $Q_e^2\to Q_m^2/2$, $Q_m^2\to 2 Q_e^2$.
However
eqs.\refb{try1}, \refb{descr} can be reexpressed in the form:
\be{try2}
\pmatrix{Q_e\cr Q_m} \to \pmatrix{d & c/2\cr 2b & a}
\pmatrix{Q_e\cr Q_m}\, ,
\ee
with $\pmatrix{d & c/2\cr 2b & a}\in \Gamma_1(2)$. This is the
usual form of S-duality transformation in the second description of
the system.

\subsection{Location of the zeroes of $\wt\Phi$}

We can follow the procedure of \cite{0605210} to identify the 
location of the zeroes of $\wt\Phi$ by examining the location
of the singularities in the integral $\II$. As in \cite{0605210} one
finds that
$\wt\Phi(\wrh ,\ws,\wv )$ has possible zeroes at
\bea{ev4.a}
&&  \left( n_2 ( \ws\wrh -\wv ^2) + b\wv   
+ n_1 \ws  - \wrh  m_1 + m_2
\right)= 0\nonumber \\
&& \quad  \hbox{for $m_1,m_2,n_2\in \ZZZ$, 
$n_1\in{1\over 2}\ZZZ$,
$b\in 2\ZZZ+1$}, \quad m_1 n_1 + m_2 n_2 
+ {b^2\over 4}={1\over 4}\,  .
\een
The order of the zero is given by
\be{eor1}
\sum_{s=0}^1 (-1)^{m_1 s} c^{(r,s)}(-1), \qquad 
\hbox{$r=2n_1$ mod 1}\, .
\ee
Using \refb{valcs} we see that \refb{eor1} vanishes
 for $r=1$. Thus
in order to get a zero (or pole), $n_1$ must be an integer.
Setting $r=0$ in \refb{eor1} and using \refb{valcs} we see that
the order of the zero is now given by $2\times (-1)^{m_1}$.
Thus $\wt\Phi$ has
second order zeroes at 
\bea{ev4}
&&  \left( n_2 ( \ws\wrh -\wv ^2) + b\wv   
+ n_1 \ws  - \wrh  m_1 + m_2
\right)= 0\nonumber \\
&& \quad  \hbox{for $m_1\in 2\ZZZ$, 
$m_2, n_2\in \ZZZ$, $n_1\in\ZZZ$,
$b\in 2\ZZZ+1$}, \quad m_1 n_1 + m_2 n_2 
+ {b^2\over 4}={1\over 4}\,  , \nonumber \\
\een
and second order poles at
\bea{ev4pole}
&&  \left( n_2 ( \ws\wrh -\wv ^2) + b\wv   
+ n_1 \ws  - \wrh  m_1 + m_2
\right)= 0\nonumber \\
&& \quad  \hbox{for $m_1\in 2\ZZZ$+1, 
$m_2, n_2\in \ZZZ$, $n_1\in\ZZZ$,
$b\in 2\ZZZ+1$}, \quad m_1 n_1 + m_2 n_2 
+ {b^2\over 4}={1\over 4}\,  . \nonumber \\
\een
We shall now determine the constant of proportionality for
two particular cases, namely near $\wv =0$ and near 
$\ws \wrh -\wv ^2+\wv =0$. The $\wv \to 0$ behaviour 
of $\wt\Phi$ can be
derived
directly from \refb{edefwphi} and the relations
\be{ecx1}
\sum_b c^{(r,s)}(4n - b^2) = \cases{0
\quad \hbox{for}\quad 
(r,s)=(0,0)\cr 8 \, \delta_{n,0} \, \, \, \quad \hbox{for}\quad
(r,s)\ne (0,0)}\, ,
\ee
which follow from setting $z=0$ in eqs.\refb{valind} and
\refb{e1.5copy}. This gives
\be{limitv}
\wt\Phi(\wrh ,\ws ,\wv )\simeq  {\pi^2\over 64}\, \wv ^2 
\frac{\eta(2\wrh )^{16} }{\eta(\wrh )^8}  \frac{\eta(\ws /2)^{16}}
{\eta(\ws )^8} \, .
\ee

In order to find the behaviour of $\wt\Phi$  
near $\ws \wrh -\wv ^2+\wv =0$ we first note from \refb{defpi10new}
 that for $v \to 0$
\be{phlim}
\Phi(\rho ,\sigma ,v ) \simeq 4\pi^2 v ^2 
\frac{\eta(2\rho )^{16} }{\eta(\rho )^8}  \frac{\eta(2\sigma )^{16}}
{\eta(\sigma )^8} + \OO(v^4)\, .
\ee
Next we use the duality transformation property
\be{ephi2}
\Phi(\rho , \sigma +2v +\rho , v  + \rho ) =\Phi(\rho , \sigma , v )\, ,
\ee
which follows from the symmetry of $\II'$ under a relabelling
of the indices $b$, $\vec m$, $\vec n$ in eq.\refb{defirslnew}.
Eqs.\refb{reln} and \refb{ephi2} give
\be{ephi3}
\wt\Phi(\wrh ,\ws ,\wv ) = \ws ^{-2} \, 
\Phi\left(\wrh  - {\wv ^2\over \ws} , {\wrh \ws -(\wv -1)^2\over
\ws },  {\ws \wrh -\wv ^2+\wv \over \ws }\right)\, .
\ee
\refb{phlim} now gives, for small 
$\wt\rho \wt\sigma - \wt v^2 + \wt v$,
 \be{e3}
\wt \Phi(\wt\rho,\wt\sigma,\wt v) = 4\pi^2\, (2v -\rho-\sigma)^2
\, v^2\, f (\rho) f (\sigma) + \OO(v^4)\, ,
\ee
where
\be{e4}
f(\rho) = \eta(2\rho)^{16} / \eta(\rho)^8\, ,
\ee
and
\be{e5}
   \rho = {\wt \rho \wt\sigma - \wt v^2\over \wt\sigma}, 
   \qquad \sigma = {\wt\rho \wt \sigma - (\wt v - 1)^2\over \wt 
   \sigma}
   \, , \qquad
   v = {\wt\rho \wt\sigma - \wt v^2 + \wt v\over \wt\sigma}\, ,
\ee
or equivalently,
\be{e6}
   \wt\rho = {v^2-\rho\sigma \over 2v-\rho-\sigma}, \qquad
   \wt\sigma={1\over 2v-\rho-\sigma}, \qquad \wt v =
   {v-\rho \over 2v-\rho-\sigma}\, .
\ee
These relations will be useful in section \ref{s1} for evaluating the
statistical entropy of the black hole.

\sectiono{Statistical  and Black Hole Entropy Functions} \label{s1}

In this section we shall compute the statistical entropy 
function\cite{0605210} of
the dyons carrying electric charge $Q_e$ and magnetic charge $Q_m$.
The value of this function at its extremum
gives the statistical entropy, -- the logarithm of the degeneracy of states
corresponding to a given set of charges. We also compute the
black hole entropy function\cite{0506177,0508042} 
whose value at its extremum
gives the Wald entropy of the black hole. 
We then compare the two results.

\subsection{Statistical entropy function} \label{s1.1}

We begin with the formula \refb{e1pre} for the
degeneracy of dyons:
\be{e1}
d(Q_e, Q_m) =   K
\int_{C} d\wt \rho d\wt\sigma d\wt v
{1\over \wt\Phi(\wt \rho, \wt \sigma, \wt v)}
\exp\left[ -i\pi (2\wt \rho Q_e^2 + \wt \sigma Q_m^2/2 +
2\wt v Q_e\cdot Q_m)\right]\, . 
\ee
This formula is identical in form to eq.(3.29) of \cite{0605210}
with the substitution $Q_m^2 \to 2 Q_e^2$, $Q_e^2\to Q_m^2/2$.
Following \cite{9607026,0605210} 
one can show that the dominant
contribution to this integral comes from
the residue at the pole at
\be{e2}
 \ws \wrh  - \wv ^2 +\wv  = 0\, .
 \ee
The behaviour of $\wt\Phi$ near this zero, given by \refb{e3}, is
identical to the corresponding relation (4.17) in
\cite{0605210} with $k\to 2$ and 
$f^{(k)}(\rho)\to f(\rho)$. Thus following an 
analysis identical to that in
\cite{0605210} we can conclude that for large charges the
statistical entropy $S_{stat}(Q_e, Q_m)$, defined as the logarithm
of the degeneracy $d(Q_e,Q_m)$, 
is obtained by extremizing the statistical entropy
function
\be{e7k}
-\wt\Gamma_B(\vec\tau') = {\pi\over 2 \tau'_{2}} \, |
{1\over \sqrt 2}   Q_m + \sqrt 2 \, \tau'\, 
Q_e |^2
- \ln f (\tau') -\ln f(-\bar\tau')
- 4 \ln (2\tau'_{2}) + \hbox{constant} + \OO(Q^{-2})\, .
\ee
with respect to the real and imaginary parts of $\tau'$.
In terms of a new variable
\be{e8}
\dum={1\over 2\bar\tau'}\, ,
\ee
we can express \refb{e7k} as
\be{e9}
-\wt\Gamma_B = {\pi\over 2\dum_{2}} \, |Q_e+\dum Q_m|^2
- \ln \wt f (\dum) -\ln \wt f(-\bar\dum)
- 4 \ln (2\dum_{2}) + \hbox{constant} + \OO(Q^{-2})\, ,
\ee
where
\be{e10}
\wt f(\dum) = \eta(\dum)^{16} / \eta(2\dum)^8\, .
\ee
For large charges the first term on the right hand side of \refb{e9}
gives the leading contribution to the statistical entropy. This
term is universal and coincides {\it e.g.} with the corresponding
term in the statistical entropy function for CHL models. The rest
of the terms, giving correction of order $Q^0$ or lower to the
entropy, depend on the specific theory being analyzed.

\subsection{Black hole entropy function} \label{s1.2}

As discussed at the end of section \ref{s0}, 
the low energy effective field theory describing the theory under
consideration is $\NN=4$ supergravity coupled to six matter
multiplets.  Since the
supergravity action  is insensitive to the details of the
theory except for the rank of the gauge group, 
the Bekenstein-Hawking entropy of
a BPS black hole carrying charges $(Q_e,Q_m)$, 
computed using
the supergravity action, reproduces the leading contribution of order
$Q^2$ to the
statistical entropy as in the case of toroidally compactified
heterotic string theory or CHL models.  However
since we shall be interested in computing the
entropy to order $Q^0$ we must also include four derivative 
corrections
to the supergravity action. An important set of four derivative
terms relevant for computing the order $Q^0$ corrections to the
entropy is the Gauss-Bonnet term. For definiteness we shall use the
second description of the theory to describe these corrections. 
On general grounds the Gauss-Bonnet term
can be shown to have the following 
structure\footnote{There is also a term
proportional to the imaginary part of the function $g(a+iS)$
multiplying the Pontryagin density. But this term does not play
any role in the analysis of the entropy of spherically symmetric
black holes since its contribution to the black hole entropy function
vanishes.}
\be{es2.1}
\Delta\LL =  \phi(a, S)\,
\left\{ R_{\mu\nu\rho\sigma} R^{\mu\nu\rho\sigma}
- 4 R_{\mu\nu} R^{\mu\nu}
+ R^2
\right\} \, ,
\ee
where $R_{\mu\nu\rho\sigma}$, $R_{\mu\nu}$ and $R$ are 
the Riemann tensor, Ricci tensor and scalar curvatures respectively,
$S=e^{-2\Phi}$ where $\Phi$ is the dilaton field and $a$ is the axion
field obtained by dualizing the rank two
anti-symmetric tensor field in four
dimensions. The function $\phi(a,S)$ 
 has the structure:
\be{es2.2}
\phi(a,S) = -{1\over 128\pi^2} \left(\KK \ln (2\, S) 
+ g(a+iS) + g(a+iS)^*\right)
\ee
where $\KK$ is a constant representing the effect of
holomorphic anomaly\cite{9302103,9309140}, 
and $g(\dum)$ is a holomorphic function of $\tau$
which will be determined shortly.
Explicit result for $\phi(a,S)$ for this model can be found in
\cite{9708062}, but we shall describe an alternative method for
determining $\phi(a,S)$ following
 \cite{0502126} which can be easily generalized to
the case of $\ZZZ_3$ orbifold to be discussed in section \ref{sz3}.
$\phi(a,S)$ is invariant under
the S-duality group $\Gamma_1(2)$, 
which acts on $\dum\equiv a+iS
\equiv \dum_1+i\dum_2$ as
\be{es2.3}
\dum\to {a\dum+b\over c\dum+d}, \quad a,b,c,d\in\ZZZ, \quad
ad-bc=1, \quad c=\hbox{0 mod 2}, \quad a,d=\hbox{1 mod 2}\, .
\ee
Thus gives
\be{es2.4}
g\left(a\dum+b\over c\dum+d\right) = g(\dum) + \KK \ln (c\dum+d)\, ,
\ee
and hence
\be{es2.5}
g(\dum) - {2\, \KK  } \ln\eta(\dum) 
\ee
is invariant under a modular transformation except for a constant shift
originating from the phases picked up by $\eta(\dum)$ under
a modular transformation.
Thus
\be{es2.6}
\p_{\dum} \left( g(\dum) - {2\, \KK } \ln\eta(\dum) \right)
\ee
must be a modular form of $\Gamma_1(2)$ of weight 2. There is a
unique modular form with this property\cite{schoen}, namely
\be{es2.7}
\p_{\dum} \left( \ln \eta(2\dum) - \ln \eta(\dum)\right)\, .
\ee
Thus \refb{es2.6} must be proportional to \refb{es2.7}. The constant
of proportionality may be determined as follows. Since toroidally
compactified type II string
theory has no Gauss-Bonnet term at the tree level, such terms are absent
even after taking the orbifold projection. This shows that $\phi(a,S)$,
and hence $g(a+iS)$
cannot have a term growing linearly with $S$ for large $S$. Comparing
the large $S$ behaviour of \refb{es2.6} and \refb{es2.7} we now
get
\be{es2.8}
g(\dum) - {2\, \KK } \ln \eta(\dum) = -{2\, \KK } 
\left( \ln \eta(2\dum) - \ln \eta(\dum)\right) +\hbox{constant}\, ,
\ee
or equivalently
\be{es2.9}
g(\dum) = -{2\, \KK } \left( \ln\eta(2\dum) - 2\ln\eta(\dum)\right)
+\hbox{constant}\, .
\ee
This gives
\bea{es2.9a}
\phi(a,S) &=& -{\KK \over 128\pi^2} \left(\ln (2\, \dum_2)
-{2} \left( \ln\eta(2\dum) - 2\ln\eta(\dum)\right) 
-{2} \left( \ln\eta(-2\bar\dum) - 2\ln\eta(-\bar\dum)\right)  \right)
\nonumber \\
&&
+\hbox{constant}\, .
\eea

Finally we turn to the determination of $\KK $. This is done following
the procedure described in \cite{0502126} with K3 replaced by $T^4$.
The net result is that $\KK $ is the number of harmonic $p$ forms on $T^4$
invariant under the transformation $\wt g$,
weighted by $(-1)^p$. Since only the even forms are invariant under
$\wt g$, and there are altogether 8 even forms on $T^4$, we get
\be{es2.10}
\KK =8\, .
\ee
This determines the structure of the Gauss-Bonnet term completely.
The result agrees with the result of explicit computation described
in 
\cite{9708062}. 

The effect of the term given in \refb{es2.1} 
on the computation of black hole entropy was
analyzed in detail in \cite{0508042}. After elimination of all variables
except the values of $a$ and $S$ on the horizon, the black hole
entropy function takes the form:
\bea{es2.11}
\EE &=& {\pi\over 2\dum_{2}} \, |Q_e+\dum Q_m|^2
+ 64\, \pi^2\, \phi(\dum_1,\dum_2)\nonumber \\
&=& {\pi\over 2\dum_{2}} \, |Q_e+\dum Q_m|^2 
-4\ln (2\dum_2) 
+{8} \left( \ln\eta(2\dum) - 2\ln\eta(\dum)\right) 
+{8} \left( \ln\eta(-2\bar\dum) - 2\ln\eta(-\bar\dum)\right)
\nonumber \\
&& +\hbox{constant}\, .
\een
Extremization of this function with respect to $\dum_1$ and 
$\dum_2$ gives the black hole entropy. Comparing 
\refb{e9} and \refb{es2.11}
 we see that the black hole entropy and the statistical
entropy agree to this
order.\footnote{We should remind the reader that the string theory
effective action has other four derivative terms besides the one
given in \refb{es2.1} and hence regarding \refb{es2.11} as the
complete contribution to the black hole entropy function to this
order is not completely justified. A somewhat different set of
four derivative terms, based on supersymmetrization of the 
curvature squared terms, give the same answer for the black hole
entropy\cite{9906094,0007195}. 
Thus it seems that the answer for the black hole entropy,
obtained by extremizing \refb{es2.11}, is somewhat robust.
Nevertheless it will be useful to determine the complete set of
four derivative corrections to the supergravity action and study
their effect on the black hole entropy. An attempt towards this
has recently been made in \cite{0607094}.
}

Given that for this model the black hole entropy fails to agree with
the statistical entropy for elementary 
string states\cite{0504005,0507014},
it is worth trying to
understand the difference between these two cases. First we note
that if we take $Q_e^2>> Q_m^2, 
(Q_e\cdot Q_m)^2/Q_m^2$ in the
expression
\refb{es2.11} for the black hole entropy function,
then extremization of the first term requires $\dum_2$ to be large.
In this limit the term growing linearly with $\dum_2$ in the rest of
the terms cancel. This does not
happen for the corresponding expression ((4.41) in
\cite{0605210}) for the
black hole entropy function for the CHL models.
Thus although the leading contribution to the black
hole entropy is the same in all $\NN=4$ supersymmetric 
compactifications, the correction to this leading term is smaller in
the present model compared to  the CHL models by powers of 
$Q_m^2/Q_e^2$ and $(Q_e\cdot Q_m)^2/Q_e^2Q_m^2$. 
This of course is a consequence of the absence of tree level
curvature squared
corrections in type II string theory.

How does this difference 
come about in the formula for the statistical entropy? 
For this we need to understand the origin of the corrections linear
in $\dum$ in the statistical entropy function. Let us for 
definiteness work in the second description of the model where
purely
electrically charged states represent elementary string states.
On physical grounds
we should expect
that when the electric charge is large compared to the magnetic
charge the correction to the leading contribution to the statistical
entropy will be dominated by the growth in the degeneracy of
elementary string states, \i.e.\ the contribution
\refb{e3.2} in the present model or its analog  in the case of
CHL models (eq.(3.2)
of \cite{0605210}). 
This intuition can be put on a firmer ground
by noting that if we remove this factor from the
dyon partition function then the modified statistical entropy,
computed using this modified partition function,  does not
contain any term growing linearly with $\tau$, either in
the present model or in the CHL models.
Thus the term in the entropy function growing linearly with
$\tau$ has its origin in the partition function of elementary
string states, and the difference in the behaviour of the statistical
entropy function in the present model and the CHL models
can be attributed to a difference in behaviour of
the elementary string partition function in the two theories.

By examining carefully the
analysis of \cite{0605210} leading to the final expression for the
statistical
entropy function one can check that the large $\dum$ behaviour of the
correction term is controlled by the small $\wrh$ behaviour of the
partition function \refb{e3.2} of elementary string states. In particular
the absence of linear corrections to the statistical entropy function of
the present model is related to the absence of exponential divergence
of \refb{e3.2} in the $\wrh\to 0$ limit. In contrast the corresponding
elementary string partition function for (say) the $\ZZZ_2$ CHL model
has the form\cite{0605210}:
\be{echel}
\eta(\wt\rho)^{-8} \eta(2\wt\rho)^{-8}\, ,
\ee
and diverges exponentially as $\wt\rho\to 0$. This difference in
behaviour might seem a bit
surprising at the first sight since the small $\wt\rho$ behaviour of
the partition function controls the growth of degeneracy for large
charges and for both models the degeneracy grows exponentially.
The difference however comes from the fact that \refb{e3.2} and
\refb{echel} actually represent an index where we multiply
the degeneracy by $(-1)^{F_L}$, $F_L$ being 
the space-time fermion
number associated with left-moving world-sheet excitations. For
the CHL model all the left-moving
excitations are bosonic and hence the degeneracy
is equal to this index. 
The exponential growth in the degeneracy causes an exponential
divergence in the partition function \refb{echel} as $\wrh\to 0$.
However for the present model, states with even
and odd momentum along $S^1$ correspond to bosonic and fermionic
states respectively\cite{0504005,0507014}, and the index is equal to
the degeneracy up to a sign. 
The small $\wrh$ behaviour of the `partition
function' \refb{e3.2} 
is controlled by the difference in the growth rate between
bosonic and fermionic excitations and the leading exponential term
cancels between these set of states. As a result \refb{e3.2} 
has no exponential divergence in the $\wrh\to 0$ limit.

To summarize the situation, we have seen that the 
absence/presence of linearly
growing correction to the statistical entropy function 
in the present/CHL model can be attributed
to the fact that in the present model elementary string spectrum contains
both bosonic and fermionic excitations in the left-moving sector, 
whereas the  CHL model
has only bosonic excitations in the left-moving sector. Nevertheless this
by itself would not provide a complete physical explanation of the
difference in behaviour of the statistical
entropy functions in the two theories 
since the statistical entropy is computed for a fixed
charge, and the elementary string states with bosonic and fermionic
left-moving excitations carry different charges.\footnote{This in 
fact is the reason
why, just as in CHL models, 
the statistical entropy of an elementary string state still grows
exponentially in this theory in disagreement with the black hole
entropy.} We must recall however that the complete
description of a state of the dyon 
involves a tensor product of states from three
different Hilbert spaces. Thus for example 
a fermionic elementary string state carrying odd momentum
along $S^1$  combined with an odd momentum state from 
another sector and a bosonic elementary string state carrying even
momentum along $S^1$, combined with an even momentum state
from another sector, 
can give rise to states carrying the same charge but opposite
statistics. Their net
contribution to the index will then be zero, causing a suppression
in the statistical entropy function. Such cancellations
will not take place in the corresponding CHL models.

This seems to be the physical explanation for
why for dyonic states
the linearly growing corrections to the statistical entropy function
are absent in the present model in agreement with the black hole
entropy, while for the statistical entropy of
elementary string states there are no such cancellations between
bosonic and fermionic states.

\sectiono{The $\ZZZ_3$ Orbifold Example} \label{sz3}

In this section we shall analyze the dyon spectrum
in another $\NN=4$ supersymmetric
theory, obtained by taking  a $\ZZZ_3$ orbifold of 
type IIA string theory compactified on a six torus $T^4\times S^1
\times \wt S^1$. The
orbifold group involves a  $2\pi/3$ rotation along one
two dimensional plane in $T^4$, $-2\pi/3$ rotation along an
orthogonal two dimensional plane in $T^4$ and $1/3$ 
unit of shift along the circle $S^1$. This of 
course requires that the $\ZZZ_3$ transformation
is a symmetry of the original torus $T^4$, -- this can be
achieved for example by taking $T^4$ to be a product of two
two dimensional tori, each with a hexagonal symmetry.
There is a dual description of these models, also as
orbifolds of type IIA string theory on a six torus
$\wh T^4\times S^1\times \wh S^1$, but now the orbifold
group involves $1/3$ unit of shift along $S^1$ 
together with a rotation by $4\pi/3$ in a coordinate plane in
$\wh T^4$
acting only on the left-moving world-sheet fields\cite{9508064}. 
As in the case of $\ZZZ_2$ orbifold model, this theory also
has $\NN=4$ supersymmetry in four dimensions. The gauge group
now has rank 10 since (in the NSR formulation)
besides all the RR sector gauge fields,
two of the gauge fields originating in the NS-NS sector are also
projected out in the second description. The S-duality group
in the second description is $\Gamma_1(3)$, consisting of
matrices $\pmatrix{a & b\cr c & d}$ satisfying
\be{descraalt}
ad-bc=1, \quad a,d\in 1 + 3\ZZZ, \quad c \in 3\ZZZ, \quad b\in \ZZZ\, .
\ee

The various parts of the analysis done for the $\ZZZ_2$ orbifold
model can be easily generalized to the case of this $\ZZZ_3$
orbifold model by following \cite{0602254,0605210}. 
For the sake of brevity we shall not repeat the
analysis here but only give the final results. Also in order to make
the comparison between the $\ZZZ_2$ and $\ZZZ_3$ models 
easier we shall state the results for $\ZZZ_N$ model which will be
valid both for $N=2$ and $N=3$. Thus by setting $N=2$ we can
recover the results of the previous sections.

First of all we note that in both models the rank $r$ of the gauge
group may be expressed as
\be{defr}
r = 2k+8\, ,
\ee
where
\be{defkk}
k +2= {12\over N+1}\, .
 \ee
The degeneracy formula takes the form
\bea{e1alt}
d(Q_e, Q_m) &=&   K
\int_{C} d\wt \rho d\wt\sigma d\wt v
{1\over \wt\Phi(\wt \rho, \wt \sigma, \wt v)}
\exp\left[ -i\pi (N\wt \rho Q_e^2 + \wt \sigma Q_m^2/N +
2\wt v Q_e\cdot Q_m)\right]\, , \nonumber \\
\een
where 
\be{edefkknew}
K = -N^{-1 - N (k+2)/(N-1)}\, ,
\ee
$C$ is  
the hypersurface
\bea{ep2alt}
Im\, \wt \rho=M_1, \quad Im \, \wt\sigma = M_2, \quad
Im \, \wt v = M_3, \nonumber \\
 0\le Re\, \wt\rho\le 1, \quad
0\le Re\, \wt\sigma\le N, \quad 0\le Re\,  \wt v\le 1\, ,
\een
and
\bea{edefwphialt}
\wt \Phi(\wt \rho,\wt \sigma,\wt v ) &=& -\, N^{-N (k+2)/(N-1)}\, 
e^{2\pi i (\wt\rho
+ \wt v)} \nonumber \\
&& \times
 \prod_{r=0}^{N-1}
\prod_{k'\in \zzz+{r\over N},l,j\in \zzz\atop k',l\ge 0, j<0 \, {\rm for}
\, k'=l=0}
\left( 1 - e^{2\pi i (\wt \sigma k'   + \wt \rho l + \wt v j)}\right)^{
\sum_{s=0}^{N-1} e^{-2\pi i sl/N } c^{(r,s)}(4lk' - j^2)}\, .
\nonumber \\
 \eea
The coefficients $c^{(r,s)}(4lk'-j^2)$ are 
 given as follows. Let us define
 \be{esi4aalt}
F^{(r,s)}(\tau,z) \equiv {1\over N} Tr_{RR;\wt g^r} \left(\wt g^s
(-1)^{F_L+F_R}
e^{2\pi i \tau L_0} e^{2\pi i {\cal J} z}\right), \qquad r,s=0,1,\cdots
N-1\, ,
 \ee
where $\wt g$ denotes the part of the orbifold action in the first
description
that acts as rotation by angles $(2\pi/N, -2\pi/N)$ on the two
orthogonal planes of
a four torus $T^4$, and
the trace is taken over all the RR sector 
states twisted by $\wt g^r$ in the $\ZZZ_N$ orbifold of
the (4,4) superconformal
field theory with target space $T^4$, --
with $\ZZZ_N$
generated by $\wt g$,  -- before we project on to $\wt g$ invariant
states. $F_L$ and $F_R$ denote the world-sheet fermion numbers 
associated with left and right chiral fermions in this SCFT, and 
${\cal J}/2$ is the  generator of the $U(1)_L$ subgroup of the
$SU(2)_L\times SU(2)_R$
R-symmetry group of this conformal field 
theory. One finds that $F^{(r,s)}(\tau,z)$ has expansion of the form
\be{enewalt}
F^{(r,s)}(\tau,z) =\sum_{b\in\zzz, n\in \zzz/N} 
c^{(r,s)}(4n -b^2)
e^{2\pi i n\tau + 2\pi i bz}\, .
\ee
This defines the coefficients $c^{(r,s)}(4n-b^2)$.

The explicit forms of $F^{(r,s)}(\tau,z)$ are as follows
\bea{valindalt}
F^{(0,s)}(\tau, z) &=& {16\over N}\, \sin^4\left(
{\pi s\over N}\right)\,  {\vt_1\left(\tau, z+{s\over N}
\right)\vt_1\left(\tau, -z+{s\over N}
\right)\over \vt_1\left(\tau,{s\over N}\right)^2} \nonumber \\
F^{(r,s)}(\tau,z) &=& {4\, N\over(N-1)^2}
\, {\vt_1\left(\tau, z+{s\over N}
+{r\over N}\tau
\right)\vt_1\left(\tau, -z+{s\over N}+{r\over N}\tau
\right)\over \vt_1\left(\tau,
{s\over N}+{r\over N}\tau\right)^2} \, ,
\nonumber \\
&& \qquad \hbox{for $1\le r\le N-1$,  $0\le s\le N-1$}
\, . 
\eea
A factor of $4\, \sin^2\left(
{\pi s\over N}\right)$ in the expression for $F^{(0,s)}(\tau,z)$
comes from the contribution due to the  right-moving
fermionic zero modes. A factor of ${4\, N^2\over(N-1)^2}$ in
the expression for $F^{(r,s)}$ counts the number of twisted sectors.
Using standard identities involving Jacobi $\vt$-functions we may
rewrite \refb{valindalt} as
\be{frsexp}
F^{(r,s)}(\tau, z) = h_0^{(r,s)}(\tau) \vt_3(2\tau, 2z) + h_1^{(r,s)}(\tau)
\vt_2(2\tau, 2z)\, ,
\ee
where
\bea{hrsexp}
h_0^{(0,s)}(\tau)&=& -{16\over N}\, \sin^4{\pi s\over N}\, {1\over
\vt_1\left(\tau, {s\over N}\right)^2}\, \vt_2\left(2\tau, 2{s\over N}
\right)\, , \nonumber \\
h_1^{(0,s)}(\tau)&=& {16\over N}\, \sin^4{\pi s\over N}\, {1\over
\vt_1\left(\tau, {s\over N}\right)^2}\, \vt_3\left(2\tau, 2{s\over N}
\right)\, , \nonumber \\
h_0^{(r,s)}(\tau)&=& -{4N\over (N-1)^2}\,  {1\over
\vt_1\left(\tau, {1\over N}(s+r\tau)\right)^2}\, \vt_2\left(2\tau, {2
\over N}(s
+r\tau)
\right)\, , \nonumber \\
h_1^{(r,s)}(\tau)&=& {4N\over (N-1)^2}\,  {1\over
\vt_1\left(\tau, {1\over N}(s+r\tau)\right)^2}\, \vt_3\left(2\tau, {2
\over N}(s
+r\tau)
\right)\, , \nonumber \\
&& \quad 0\le s\le (N-1), \quad 1\le r\le (N-1)\, .
\eea
The coefficients $c^{(r,s)}(u)$ may now be defined through the
expansion
\be{ecrsfin}
h^{(r,s)}_l(\tau)= \sum_{n\in {1\over N}\zzz-{l\over 4}}
c^{(r,s)}(4n) e^{2\pi i n\tau}\, .
\ee
{}From \refb{ecrsfin} 
one can calculate the coefficients $c^{(r,s)}(u)$ 
explicitly.\footnote{Incidentally, the coefficients $c^{(r,s)}(u)$ are
related to the corresponding coefficients  
for the $\ZZZ_N$ CHL model (which we shall
denote by $c^{(r,s)}_{chl}(u)$)
via the relations
\be{enewre}
c^{(r,s)}(u) = \cases{ 0 \quad {\rm for} \quad (r,s)=(0,0)\cr
{N\over N-1}\, c^{(r,s)}_{chl}(u) \quad {\rm for} \quad (r,s)\ne
(0,0)\, .}
\ee
}

Generalizing the analysis of section \ref{s3} one can show that
the function $\wt\Phi(\wt \rho,\wt \sigma,\wt v)$ transforms
as a
modular form of weight $k$ under a certain subgroup of the Siegel
modular group of genus two Riemann surfaces, with
$k$ given by \refb{defkk}. 
This subgroup is the same one that appears in the analysis of
\cite{0510147,0602254} for
$\ZZZ_N$ CHL model.
Using this fact one
can prove that the degeneracy formula \refb{e1alt} is invariant under
the S-duality group $\Gamma_1(N)$ of the theory. 

Analysis of the behaviour of the statistical entropy for large charges
shows that this is given by extremizing a statistical entropy function
\be{blackalt}
{\pi\over 2\dum_{2}} \, |Q_e+\dum Q_m|^2
- \ln \wt f_k (\dum) -\ln \wt f_k(-\bar\dum)
- (k+2) \ln (2\dum_{2}) + \hbox{constant} + \OO(Q^{-2})\, ,
\ee
with respect to the real and imaginary parts of $\dum=\dum_1
+i\dum_2$. Here
\be{e10alt}
\wt f_k(\dum) = \eta(\dum)^{2N(k+2)/(N-1)}
\eta(N\dum)^{-2(k+2)/(N-1)}\, .
\ee
In order to compute the black hole entropy we need to determine the
function $\phi(a,S)$ introduced in \refb{es2.1}.
This can be done by generalizing the analysis of section
\ref{s1.2}; all that changes is that \refb{es2.7} now takes the
form $\p_{\dum} \left( \ln \eta(N\dum) - \ln \eta(\dum)\right)$
and $\KK$ is given by $2k+4$. The result is
\bea{altdefphi}
\phi(a,S) &=& -{k+2 \over 64\pi^2} \bigg(\ln (2\, \dum_2)
-{2\over N-1} \left( \ln\eta(N\dum) - N
\ln\eta(\dum)\right) \nonumber \\
&&
-{2\over N-1} \left( \ln\eta(-N\bar\dum) - 
N\ln\eta(-\bar\dum)\right)  \bigg) +\hbox{constant}
\eea
Using this the black hole entropy function becomes
\be{blackalt1}
{\pi\over 2\dum_{2}} \, |Q_e+\dum Q_m|^2
- \ln \wt f_k (\dum) -\ln \wt f_k(-\bar\dum)
- (k+2) \ln (2\dum_{2}) + \hbox{constant}  \, .
\ee
Thus again we see that the black hole entropy and the statistical
entropy match to this order.

Finally, to complete the comparison with the corresponding analysis
for the CHL orbifold models, we  note that it is possible
to find 
a series formula for the modular form $\wt\Phi$ 
and its closely related
cousin $\Phi$ defined through
\be{ephiwtphiapp}
\wt\Phi(\wrh ,\ws,\wv ) = -(-i\ws)^{-k}\,  \Phi\left(
\wrh  -{\wv ^2\over \ws},-{1\over \ws},
     {\wv \over \ws}\right)\, ,
\ee
by repeating 
the analysis of 
\cite{0510147}. This is done by
replacing the cusp form $f^{(k)}(\tau)$ used
in \cite{0510147} by the modular form
\be{enewmod}
f_k(\tau)=\eta(N\tau)^{2 N (k+2)/(N-1)}
\eta(\tau)^{-2(k+2)/(N-1)}
\ee
of $\Gamma_1(N)$ of weight $(k+2)$.
Both for $N=2$ and $N=3$, $f_k(\tau)$ vanishes as $q=e^{2\pi i\tau}$
at the cusp at $\tau\to i\infty$. However using the modular transformation
properties of $\eta(\tau)$ it is easy to see that $\tau^{-k-2} f_k(-1/\tau)$
goes to a constant as $\tau\to i\infty$. Thus $f_k(\tau)$ is not a cusp
form of $\Gamma_1(N)$. Nevertheless 
we can proceed as in  \cite{0510147} to construct (meromorphic)
modular forms $\Phi$ and $\wt\Phi$ of weight $k$
of appropriate subgroups
of $Sp(2,\ZZZ)$. For example the modular form $\Phi$ is
given by a formula analogous to eq.(1.6) of
\cite{0510147}
\be{esk3a}
   \Phi(\rho,\sigma, v)  = \sum_{n,m,r\in\zzz\atop
     n,m\ge 1, \, r^2< 4mn}\, a(n,m,r)
   \, e^{2\pi i (n\rho+m\sigma+rv)}
   \, ,
\ee
where,
\be{esk4a}
   a(n,m,r) = \sum_{\alpha\in \zzz;\alpha>0\atop\alpha|(n,m,r), \,
     {\small \rm g.c.d.}(\alpha,N)=1} \chi(\alpha)\,
   \alpha^{k-1}\, C\left({4mn
       -r^2\over \alpha^2}\right)\, ,
\ee
\bea{edefchi}
\chi(\alpha)&=&1\quad \hbox{for $N=2$}\cr \cr
&=&  \cases{\hbox{1 \,  \,
for $\alpha=1$ mod 3}\cr \hbox{$-1$ for
$\alpha=2$ mod 3}} \quad \hbox{for $N=3$}\, .
\eea 
The coefficients $C(m)$ are obtained from the 
modular form $f_k(\tau)$ 
as follows.  We first define the coefficients $f_{k,n}$ as
\be{esk1a}
  f_k(\tau)\eta(\tau)^{-6}
= \sum_{n\ge 1} f_{k,n} e^{2\pi i \tau(n-{1\over 4})} \, ,
\ee
and then define $C(m)$ as
\be{esa3}
   C(m) = (-1)^m \sum_{s,n\in\zzz \atop n\ge 1} f_{k,n}
   \delta_{4n+{s^2-1 },{m }}\, .
\ee
Eq.\refb{esk3a} gives a series expansion for $\Phi$.
A similar series
expansion for $\wt\Phi$ may be found by following the analysis of
\cite{0510147} (see eq.(C.37) of this paper)
but since the formul\ae\ are considerably more
complicated we shall not describe it here.

\bigskip

{\bf Acknowledgement:} We would like to thank E.~Gava and 
K.S.~Narain for useful discussions.


\begin{thebibliography}{99}

\bibitem{9607026}
R.~Dijkgraaf, E.~P.~Verlinde and H.~L.~Verlinde,
``Counting dyons in N = 4 string theory,''
Nucl.\ Phys.\ B {\bf 484}, 543 (1997)
[arXiv:hep-th/9607026].

\bibitem{0412287}
G.~L.~Cardoso, B.~de Wit, J.~Kappeli and T.~Mohaupt,
``Asymptotic degeneracy of dyonic N = 4 string states
and black hole
entropy,''
JHEP {\bf 0412}, 075 (2004) [arXiv:hep-th/0412287].

\bibitem{0505094}
D.~Shih, A.~Strominger and X.~Yin,
``Recounting dyons in N = 4 string theory,''
arXiv:hep-th/0505094.

\bibitem{0506249}
D.~Gaiotto,
``Re-recounting dyons in N = 4 string theory,''
arXiv:hep-th/0506249.

\bibitem{0508174}
D.~Shih and X.~Yin,
``Exact black hole degeneracies and the topological string,''
arXiv:hep-th/0508174.

\bibitem{0510147}
D.~P.~Jatkar and A.~Sen,
``Dyon spectrum in CHL models,''
arXiv:hep-th/0510147.

  \bibitem{0602254}
  J.~R.~David, D.~P.~Jatkar and A.~Sen,
  ``Product representation of dyon 
  partition function in CHL models,''
  arXiv:hep-th/0602254.
  
 \bibitem{0603066}
  A.~Dabholkar and S.~Nampuri,  
  ``Spectrum of dyons and black holes in 
  CHL orbifolds using Borcherds lift,''
  arXiv:hep-th/0603066.

\bibitem{0605210}
  J.~R.~David and A.~Sen,
   ``CHL dyons and statistical entropy 
   function from D1-D5 system,''
  arXiv:hep-th/0605210.
  
  \bibitem{CHL}
S.~Chaudhuri, G.~Hockney and J.~D.~Lykken,
``Maximally supersymmetric string theories in D $<$ 10,''
Phys.\ Rev.\ Lett.\  {\bf 75}, 2264 (1995)
[arXiv:hep-th/9505054].

\bibitem{CP}
S.~Chaudhuri and J.~Polchinski,
``Moduli space of CHL strings,''
Phys.\ Rev.\ D {\bf 52}, 7168 (1995)
[arXiv:hep-th/9506048].

\bibitem{9507027}
J.~H.~Schwarz and A.~Sen,
 ``Type IIA dual of the six-dimensional CHL compactification,''
Phys.\ Lett.\ B {\bf 357}, 323 (1995) [arXiv:hep-th/9507027].

\bibitem{9507050}
C.~Vafa and E.~Witten,
``Dual string pairs with N = 1 and N = 2 supersymmetry in four
dimensions,''
Nucl.\ Phys.\ Proc.\ Suppl.\  {\bf 46}, 225 (1996)
[arXiv:hep-th/9507050].

\bibitem{9508144}
S.~Chaudhuri and D.~A.~Lowe,
``Type IIA heterotic duals with maximal supersymmetry,''
Nucl.\ Phys.\ B {\bf 459}, 113 (1996)
[arXiv:hep-th/9508144].

\bibitem{9508154}
P.~S.~Aspinwall,
``Some relationships between dualities in string theory,''
Nucl.\ Phys.\ Proc.\ Suppl.\  {\bf 46}, 30 (1996)
[arXiv:hep-th/9508154].

\bibitem{9307038}
R.~M.~Wald,
``Black hole entropy in the Noether charge,''
Phys.\ Rev.\ D {\bf 48}, 3427 (1993)
[arXiv:gr-qc/9307038].

\bibitem{9312023}
 T.~Jacobson, G.~Kang and R.~C.~Myers,
 ``On black hole entropy,''
Phys.\ Rev.\ D {\bf 49}, 6587 (1994) [arXiv:gr-qc/9312023].

\bibitem{9403028}
V.~Iyer and R.~M.~Wald,
``Some properties of Noether charge and a proposal for dynamical black
hole
entropy,''
Phys.\ Rev.\ D {\bf 50}, 846 (1994)
[arXiv:gr-qc/9403028].

\bibitem{9502009}
T.~Jacobson, G.~Kang and R.~C.~Myers,
``Black hole entropy in higher curvature gravity,''
arXiv:gr-qc/9502009.

\bibitem{9508064}
  A.~Sen and C.~Vafa,
   ``Dual pairs of type II string compactification,''
  Nucl.\ Phys.\ B {\bf 455}, 165 (1995)
  [arXiv:hep-th/9508064].
  
 \bibitem{0504005}
  A.~Sen,
   ``Black holes and the spectrum of 
   half-BPS states in N = 4 supersymmetric
  string theory,''
  Adv.\ Theor.\ Math.\ Phys.\  {\bf 9}, 527 (2005)
  [arXiv:hep-th/0504005].
   
 \bibitem{0507014}
  A.~Dabholkar, F.~Denef, G.~W.~Moore and B.~Pioline,
   ``Precision counting of small black holes,''
  JHEP {\bf 0510}, 096 (2005)
  [arXiv:hep-th/0507014].
  
\bibitem{0503217}
  D.~Gaiotto, A.~Strominger and X.~Yin,
 ``New connections between 4D and 5D black holes,''
  JHEP {\bf 0602}, 024 (2006)
  [arXiv:hep-th/0503217].
  
\bibitem{0203048}
  J.~R.~David, G.~Mandal and S.~R.~Wadia,
   ``Microscopic formulation of black holes in string theory,''
  Phys.\ Rept.\  {\bf 369}, 549 (2002)
  [arXiv:hep-th/0203048].
  
\bibitem{9602065}
  J.~C.~Breckenridge, R.~C.~Myers, A.~W.~Peet and C.~Vafa,
  ``D-branes and spinning black holes,''
  Phys.\ Lett.\ B {\bf 391}, 93 (1997)
  [arXiv:hep-th/9602065].
  
\bibitem{9608096}
R.~Dijkgraaf, G.~W.~Moore, E.~P.~Verlinde and 
H.~L.~Verlinde,
``Elliptic genera of symmetric products and second quantized strings,''
  Commun.\ Math.\ Phys.\  {\bf 185}, 197 (1997)
  [arXiv:hep-th/9608096].
  
 \bibitem{dixon} 
L.~J.~Dixon, V.~Kaplunovsky and J.~Louis, 
``Moduli
Dependence Of String Loop Corrections To Gauge Coupling 
Constants,''
Nucl.\ Phys.\ B {\bf 355}, 649 (1991).

  \bibitem{9512046}
T.~Kawai,
``$N=2$ heterotic string threshold correction, $K3$ surface and 
generalized Kac-Moody superalgebra,''
Phys.\ Lett.\ B {\bf 372}, 59 (1996)
[arXiv:hep-th/9512046].


\bibitem{0506177}
  A.~Sen,
   ``Black hole entropy function and the attractor mechanism in higher
   derivative gravity,''
  JHEP {\bf 0509}, 038 (2005)
  [arXiv:hep-th/0506177].

\bibitem{0508042}
  A.~Sen,
   ``Entropy function for heterotic black holes,''
  JHEP {\bf 0603}, 008 (2006)
  [arXiv:hep-th/0508042].
  
\bibitem{9302103}
  M.~Bershadsky, S.~Cecotti, H.~Ooguri and C.~Vafa,
  ``Holomorphic anomalies in topological field theories,''
  Nucl.\ Phys.\ B {\bf 405}, 279 (1993)
  [arXiv:hep-th/9302103].
  
\bibitem{9309140}
  M.~Bershadsky, S.~Cecotti, H.~Ooguri and C.~Vafa,
   ``Kodaira-Spencer theory of gravity 
   and exact results for quantum string
  amplitudes,''
  Commun.\ Math.\ Phys.\  {\bf 165}, 311 (1994)
  [arXiv:hep-th/9309140].


\bibitem{9708062}
  A.~Gregori, E.~Kiritsis, C.~Kounnas, 
  N.~A.~Obers, P.~M.~Petropoulos and B.~Pioline,
 ``R$^2$ corrections and 
 non-perturbative dualities of N = 4 string ground
   states,''
  Nucl.\ Phys.\ B {\bf 510}, 423 (1998)
  [arXiv:hep-th/9708062].

\bibitem{0502126}
  A.~Sen,
   ``Black holes, elementary strings and holomorphic anomaly,''
  JHEP {\bf 0507}, 063 (2005)
  [arXiv:hep-th/0502126].

\bibitem{schoen}
B.~Schoeneberg, 
Elliptic Modular Functions, Springer Verlag, 1974.

\bibitem{9906094}
  G.~Lopes Cardoso, B.~de Wit and T.~Mohaupt,
   ``Macroscopic entropy formulae 
   and non-holomorphic corrections for
   supersymmetric black holes,''
   Nucl.\ Phys.\ B {\bf 567}, 87 (2000)
  [arXiv:hep-th/9906094].
  
\bibitem{0007195}
  T.~Mohaupt,
   ``Black hole entropy, special geometry and strings,''
  Fortsch.\ Phys.\  {\bf 49}, 3 (2001)
  [arXiv:hep-th/0007195].
  
\bibitem{0607094}
  G.~Exirifard,
  ``The world-sheet corrections to dyons in the Heterotic theory,''
  arXiv:hep-th/0607094.

\end{thebibliography}
 \end{document}